\documentclass[12pt,amsmath,amssymb,nofootinbib,aps,prl]{revtex4}

\usepackage{graphicx}

\begin{document}

\title{Symmetry, causal structure and superluminality\\ in Finsler spacetime}

\author{Zhe Chang}
\email{changz@ihep.ac.cn}
\author{Xin Li}
\email{lixin@ihep.ac.cn}
\author{Sai Wang}
\email{wangsai@ihep.ac.cn}

\affiliation{\scriptsize{Institute of High Energy Physics\\
and\\ Theoretical Physics Center for Science Facilities
\vspace{0.15cm}\\
Chinese Academy of Sciences, 100049 Beijing, China}}


\begin{abstract}

The superluminal behaviors of neutrinos were reported by the OPERA collaboration recently.
It was also noticed by Cohen and Glashow that, in standard quantum field theory,
the superluminal neutrinos would lose their energy via the Cherenkov-like process rapidly.
Finslerian special relativity may provide a framework to cooperate with the OPERA neutrino superluminality without Cherenkov-like process.
We present clearly the symmetry, causal structure and superluminality in Finsler spacetime.
The principle of relativity and the causal law are preserved.
The energy and momentum are well defined and conserved in Finslerian special relativity.
The Cherenkov-like process is proved to be forbidden kinematically and
the superluminal neutrinos would not lose energy in their distant propagations from CERN to the Gran Sasso Laboratory.
The energy dependence of neutrino superluminality is studied based on the reported data of the OPERA collaboration as well as other groups.

\end{abstract}

\maketitle

\section{I. Introduction}

The Lorentz invariance (LI) is one of the fundamentals of modern physics.
It is meaningful to test the fate of LI both on theories and experiments.
Kostelecky and Samuel \cite{KosteleckyS001} have manifested that the LI could be broken spontaneously in the string theory.
The spontaneous Lorentz breaking involves the expectation values of Lorentz vectors and tensors in the Lagrangian of particles
which lead to the framework of standard model extension (SME) \cite{SME}.
Coleman and Glashow \cite{ColemanGlashowLIV} have proposed a perturbative framework to investigate the possible departures from the LI,
in which the spacetime translations and space rotations are invariant while the Lorentz boosts have small departures.
In a different approach, Cohen and Glashow \cite{VSR} suggested that
the symmetry group of nature is isomorphic to the spacetime translation group plus a proper subgroup of the Lorentz group,
which is referred as the theory of very special relativity (VSR).
In addition, the Lorentz transformations were deformed in the doubly special relativity (DSR) \cite{DSR}
because of the Planckian-scale effects of quantum gravity.

Recently, one possible signal of Lorentz invariance violation (LIV) was reported by the OPERA collaboration that
the muon neutrinos behave superluminally \cite{OPERA2011}.
The muon neutrinos were produced in the CERN and arrived at the Gran Sasso Laboratory in advance
than expectation by Einstein's special relativity.
To study the energy dependence of neutrino superluminality, the data of the OPERA neutrino experiment was split into two groups.
The speed is reported as \(1+\left(2.18\pm0.77\pm0.30\right)\times10^{-5}\)
for the neutrinos with energy below \(20~\rm{GeV}\) with the mean energy \(13.9~\rm{GeV}\);
the speed is reported as \(1+\left(2.75\pm0.75\pm0.30\right)\times10^{-5}\)
for the neutrinos with energy above \(20~\rm{GeV}\) with the mean energy \(42.9~\rm{GeV}\).
Throughout of the paper, we use the natural unit which implies that \(c=1\).
The previous neutrino experiments or observations also gave evidences or constraints
on the superluminal behaviors \cite{MINOS2007,Fermilab1979,SN1987A}.

Soon after the OPERA's report, Cohen and Glashow \cite{CohenG01} pointed out that, in the framework of standard quantum field theory,
the superluminal neutrinos would lose their energy via the Cherenkov-like process (\(\nu\longrightarrow\nu+e^{-}+e^{+}\)) rapidly
in their distant propagations from the CERN to the Gran Sasso Laboratory.
The number of the superluminal muon neutrinos detected by the OPERA detector should be suppressed strongly.
The OPERA detector would not receive the neutrinos with energy above \(12.5~\rm{GeV}\)
which is contradictory with the results of the OPERA experiment.
Bi {\it et al.} \cite{BiETAL01} made a similar discussion on this issue.
Their arguments are in the context of LIV with a preferred frame.
Only in the preferred frame, the energy-momentum conservation is preserved \cite{VillanteV001,Amelino-CameliaETAL01}.
Furthermore, Li {\it et al.} \cite{LiLMWZ001} pointed that the Cherenkov-like process is unavoidable even
in the trivial frame without the effective ''rest frame''.
However, Amelino-Camelia {\it et al.} \cite{Amelino-CameliaETAL01} revealed that
the Cherenkov-like process is forbidden in the context that the principle of relativity is preserved
and the energy-momentum conservation is amended.
In addition, the ICARUS collaboration \cite{AntonelloETAL001} reported that
there are no Cherenkov-like events observed directly for the superluminal neutrinos.

The superluminality of particles is stringently forbidden in Einstein's special relativity.
The speed of light is the upper limit of speed for all particles unless in the context of LIV.
To account for the data of the neutrino superluminality, the dispersion relations are considered phenomenally
\begin{equation}
\label{Dispersion relations}
\eta^{\mu\nu}p_{\mu}p_{\nu}=m^{2}-\sum_{n=1}^{\infty}A_{n}(\mu,M)p_{0}^{n}\ ,
\end{equation}
where the \(A_{n}\) are dimensional coefficients which are functions of the physical mass scale of particles \(\mu\)
and the energy scale of new physics \(M\).
For a given nonvanishing power exponent \(n\), the superluminal neutrinos propagate with energy dependence as \(\delta v:=v-1\propto E^{n}\),
where \(E\) denotes the energy of neutrinos.
This is a power-law energy dependence for the superluminal behaviors of neutrinos,
and the power exponent \(n\) could be constrained by the neutrino observations.

If the OPERA's report is confirmed, Einstein's special relativity as well as the Minkowskian description of spacetime should be amended.
The superluminality of neutrinos may imply new spacetime structure.
In the new spacetime, the superluminality of particles at least neutrinos is admitted
and consistent with the present neutrino experiments and observations.
Meanwhile, the causality still holds and the Cherenkov-like process is forbidden.
The superluminal neutrinos could arrive at the OPERA detector from the distant CERN without losing their energy rapidly.

The Finslerian spacetime has been proposed to be a reasonable candidate to account for the
neutrino superluminality \cite{Finsler special relativity,Finslerian special relativity Cherenkov-like process forbidden}.
The Finsler geometry \cite{Book by Bao and Shen} is a straightforward generalization of the Riemann geometry
without the quadratic restriction on the metric, which may introduce new insights on the spacetime background.
The Finsler spacetime structure is dependent on one or more preferred directions.
The LIV was studied in the Finsler spacetimes with modified physical dispersion relations \cite{GirelliLS001,MDR,ChangL001}.
It is worth to note that the modified dispersion relations in the DSR
could be realized in the Finsler geometry \cite{GirelliLS001}.
The VSR \cite{VSR} was proved to reside in Finsler spacetime \cite{VSR Finsler}.
Most recently, the effective fields with LIV, namely SME, was proposed to be linked to Finsler geometry by Kostelecky \cite{KosteleckyFinslerSME}.
In addition, the symmetry of special relativity in the Finsler spacetimes with constant curvature was studied systematically \cite{LiC001}.
Furthermore, the Finsler geometry could also bring about new insights on the resolution of the anomalies residing
in Einstein's general relativity and cosmology \cite{Finsler gravity}.

We have proposed \cite{Finsler special relativity} a Finslerian special relativity of (\(\alpha,\beta\)) type
with an additional term which is three orders of \(\beta/\alpha\) in the line element of spacetime.
A preferred direction was involved in the line element of Finslerian special relativity to account for the superluminal behaviors of neutrinos.
The null structure was found to be enlarged and the causality was still preserved for superluminal neutrinos.
We studied the kinematics and obtained a new dispersion relation of the form (\ref{Dispersion relations}) with only \(A_{3}\neq0\).
Then the superluminality was found to be linearly dependent on the energy per unit mass of particles,
which is roughly consistent with the present neutrino experiments and observations.
Besides these, we proved that the energy-momentum conservation is preserved
and the energy-momentum is well defined in Finslerian special relativity \cite{Finslerian special relativity Cherenkov-like process forbidden}.
The Cherenkov-like process is forbidden for the superluminal neutrinos.
The superluminal neutrinos would not lose their energy rapidly via this Cherenkov-like process.
After a distant propagation from CERN to the Gran Sasso Laboratory, a large quantity of superluminal neutrinos survive
and could be detected by the OPERA detector.

In the present paper, we investigate the symmetry, causal structure and superluminality clearly in Finslerian special relativity.
The general case of energy dependence of superluminality is studied.
It is found that the energy and momentum conservations are preserved and the Cherenkov-like process is forbidden for the superluminal neutrinos.
The predicted energy dependence of neutrino superluminality in Finslerian special relativity are compared with
data of the superluminal neutrino experiments and astrophysical observations.
The rest of the paper is arranged as follows.
In Section II, the line element of Finslerian special relativity is proposed
and the corresponding dispersion relation is obtained.
We show clearly superluminality of particles and enlarged causal structure.
In Finslerian special relativity, the causality and the energy-momentum conservation are preserved.
The Cherenkov-like process is proved to be forbidden.
In Section III, the energy dependence of the superluminal behaviors of neutrinos is studied in the framework of Finslerian special relativity.
Discussions and remarks are listed in Section IV.

\section{II. Theory: Finslerian special relativity and superluminality}

In this section, we present Finslerian special relativity with general power-law energy dependence of neutrino superluminality.
The corresponding dispersion relations are obtained and the null structure is found to be enlarged.
The energy and momentum are found to be well defined and proved to be conserved.
The Cherenkov-like process is proved to be forbidden kinematically.

\subsection{A. Finslerian line element}

The action of free particles in Finslerian special relativity takes the form
\begin{equation}
\label{integral length}
I\propto\int F\left(x, y\right)d\tau\ ,
\end{equation}
where \(x^{\mu}\) and \(y^{\mu}:=dx^{\mu}/d\tau\) denote the position and four-velocity of particles, respectively.
The Greek indices run from \(0\) to \(3\) and the Latin indices run from \(1\) to \(3\).
The integrand \(F\) is positively homogeneous of order one \cite{Book by Bao and Shen}.
The metric tensor in the Finsler spacetime is defined by
\begin{equation}
g_{\mu\nu}:=\frac{\partial}{\partial y^\mu}\frac{\partial}{\partial y^\nu}\left(\frac{1}{2}F^2\right)\ ,
\end{equation}
which is used to lower and raise the indices of vectors and tensors together with its inverse.

The physical spacetime may be described by the Finsler structure which depart mildly from the Minkowski one.
Suppose that the Finslerian line element take the simple form
\begin{equation}
\label{Finslerian special relativity line elements}
F(y)d\tau=\alpha\left(1-A\left(\frac{\beta}{\alpha}\right)^{n+2}\right)d\tau\ ,
\end{equation}
where
\begin{eqnarray}
\alpha&=&\sqrt{\eta_{\mu\nu}y^{\mu}y^{\nu}}\ ,\\
\beta&=&b_{\mu}y^{\mu}\ ,
\end{eqnarray}
and \(\eta_{\mu\nu}\) is the Minkowski metric, \(b_{\mu}=(1,0,0,0)\) and \(n\) denotes a non-negative real number.
The dimensionless parameter \(A\) characterizes the level of the departure of Finslerian special relativity from Einstein's special relativity.
And the parameter \(A\) takes a tiny value which could be determined uniquely by the superluminal experiments and observations.
It is noted that the Finslerian line element (\ref{Finslerian special relativity line elements}) is locally Minkowskian \cite{Book by Bao and Shen}
and belongs to the (\(\alpha,\beta\)) type \cite{alpha beta type}.
Furthermore, the Minkowski line element could be added more extra terms in powers of \(\beta/\alpha\)
to generate the dispersion relation in Eq.(\ref{Dispersion relations}).

\subsection{B. Physical dispersion relations and superluminality}

For the particles with mass, the normalization of the Finsler norm is \(F(y)=1\).
The canonical four-momentum of the particle with mass \(m\) is given by
\begin{equation}
\label{four-momentum}
p_{\mu}:=m\frac{\partial F}{\partial y^{\mu}}\ ,
\end{equation}
which is a conserved quantity.
Corresponding to the Finslerian line element (\ref{Finslerian special relativity line elements}),
the kinematics implies the physical dispersion relation
\begin{equation}
g^{\mu\nu}p_{\mu}p_{\nu}=m^{2}\ ,
\end{equation}
which could be rewritten as
\begin{equation}
\label{Finslerian special relativity dispersion relation}
\eta^{\mu\nu}p_{\mu}p_{\nu}=m^{2}-2Ap_{0}^{2}\left(\frac{p_{0}}{m}\right)^{n}\ ,
\end{equation}
where we have neglected the terms with higher orders of \(A\).
This could also be demonstrated by the correspondence between the dispersion relations and the Finsler line elements \cite{GirelliLS001}.
In the case of large enough \(p_{0}\) and \(A>0\), the right hand side of equation (\ref{Finslerian special relativity dispersion relation})
is negative and the superluminal behaviors of particles emerge.
The speed of particle is defined as \cite{CacciapagliaDP001}
\begin{equation}
v:=\frac{\sqrt{-\eta^{ij}p_{i}p_{j}}}{\sqrt{\eta^{00}p_{0}p_{0}}}\approx1-\frac{1}{2u^{2}}+Au^{n}\ ,
\end{equation}
where \(u\) denotes the energy per unit mass \({E}/{m}\).
It is demonstrated that the speed of particles could be larger than one when \(A>0\) and \(u\) is large enough.

\subsection{C. Null structure and causality}

\begin{figure}[h]
\begin{center}
\includegraphics[width=8cm]{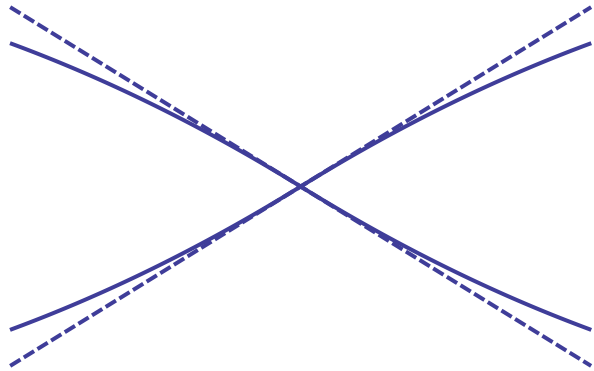}
\caption{\small{A schematic plot for the Finslerian null structure. The dashed line denotes the null structure in Einstein's special relativity
and the real line denotes the one in Finsler spacetime.
The null structure is found to be enlarged in Finslerian special relativity.}}
\label{fig1}
\end{center}
\end{figure}

To study the null structure, the Finsler norm is normalized to be \(F(y)=0\).
The causal four-velocity is defined by
\begin{equation}
u_{\mu}:=\frac{\partial F}{\partial y^{\mu}}\ .
\end{equation}
The null structure is obtained as
\begin{equation}
\eta^{\mu\nu}u^{'}_{\mu}u^{'}_{\nu}=-2A(u^{'}_{0})^{n+2}\ ,
\end{equation}
where the primes denote the normalization with respect to $F$.
It could be seen that the superluminal causal speed is admitted in this null structure when the right hand side of the above equation is negative.
The causal speed is given by
\begin{equation}
v_{c}:=\frac{\sqrt{-\eta^{ij}u^{'}_{i}u^{'}_{j}}}{\sqrt{\eta^{00}u^{'}_{0}u^{'}_{0}}}\approx1+A\left(u^{'}_{0}\right)^{n}\ .
\end{equation}
It is found that the null structure is enlarged in Finslerian special relativity than that in Einstein's special relativity when \(A>0\).
In addition, the superluminal behaviors of neutrinos would not break the causality
since the speed of neutrinos is always smaller than the causal speed.
This null structure is illustrated schematically in the Fig.\ref{fig1}.

\subsection{D. Energy-momentum conservation}

It is obvious that the Finslerian line element (\ref{Finslerian special relativity line elements}) is invariant under the spacetime translations
since it does not depend on the spacetime positions.
This could also be demonstrated by the approach of isometry or equivalently Killing vectors \cite{LiC001}.
The infinitesimal coordinate transformation is
\begin{eqnarray}
\label{infinitesimal transformation 1}
x^{\mu}&\longrightarrow& x^{\mu}+\epsilon V^{\mu}\ ,\\
\label{infinitesimal transformation 2}
y^{\mu}&\longrightarrow& y^{\mu}+\epsilon \frac{\partial V^{\mu}}{\partial x^{\nu}}y^{\nu}\ ,
\end{eqnarray}
where \(|\epsilon|\ll 1\) and the generators are called Killing vectors \(V^{\mu}\).
The Finsler structure is isometry if and only if
\begin{equation}
F(x,y)=F(\bar{x},\bar{y})\ ,
\end{equation}
under the coordinate transformation (\ref{infinitesimal transformation 1}) and (\ref{infinitesimal transformation 2}).
For the Finslerian line element (\ref{Finslerian special relativity line elements}) of (\(\alpha,\beta\)) type,
the isometry implies that the Killing vectors satisfy
\begin{equation}
\label{Killing equations}
V^{\mu}\frac{\partial F}{\partial x^{\mu}}+y^{\nu}\frac{\partial V^{\mu}}{\partial x^{\nu}}\frac{\partial F}{\partial y^{\mu}}=0\ .
\end{equation}
It is obvious that the constant vectors \(V^{\mu}=C^{\mu}\) are solutions of
the Killing equation (\ref{Killing equations}) for the Finslerian line element (\ref{Finslerian special relativity line elements}).
Based on the Noether theorem, the spacetime translational invariance implies that
the energy and momentum \(p_{\mu}\) are well defined and conserved in Finslerian special relativity.

\subsection{E. No Cherenkov-like process}

Based on the energy-momentum conservation and the dispersion relation (\ref{Finslerian special relativity dispersion relation}),
the Cherenkov-like process could be proved to be forbidden in Finslerian special relativity.
It is enough to describe properties of the Cherenkov-like process by the process (\(\mu\longrightarrow M+M\)) \cite{Amelino-CameliaETAL01}.
There are only one single incoming particle with mass \(\mu\), energy \(E\), and momentum \({P}\)
while two ejected particles with mass both \(M\), energy \(E_{1}\), \(E_{2}\),
and momentum \({P}_{1}\), \({P}_{2}\).
Meanwhile, the incoming particle is more light than the two ejected particles (\(\mu<M\)).
The energy-momentum conservation in Finslerian special relativity implies that
\begin{eqnarray}
\label{energy conservation}
E&=&E_{1}+E_{2}\ ,\\
\label{momentum conservation}
{P}^{2}&=&{P}_{1}^{2}+{P}_{2}^{2}+2P_{1}P_{2}\cos\theta\ ,
\end{eqnarray}
where \(\theta\) is the angle between the moving directions of the two ejected particles.
By combining the four-momentum (\ref{four-momentum}) and the dispersion relation (\ref{Finslerian special relativity dispersion relation})
with the energy-momentum conservation (\ref{energy conservation}) and (\ref{momentum conservation}), we obtain
\begin{eqnarray}
\cos\theta&=&\frac{2E_{1}E_{2}+2A\left(\frac{(E_{1}+E_{2})^{n+2}}{\mu^{n}}-\frac{E_{1}^{n+2}+E_{2}^{n+2}}{M^{n}}\right)
-\mu^{2}+2M^{2}}{2E_{1}E_{2}+2AE_{1}E_{2}\frac{E_{1}^{n}+E_{2}^{n}}{M^{n}}-M^{2}\left(\frac{E_{1}}{E_{2}}+\frac{E_{2}}{E_{1}}\right)}
+\mathcal{O}(A^{2})\nonumber\\
&=&1+\frac{2A\left(\frac{(E_{1}+E_{2})^{n+2}}{\mu^{n}}-\frac{E_{1}^{n+2}+E_{2}^{n+2}}{M^{n}}\right)-2AE_{1}E_{2}\frac{E_{1}^{n}+E_{2}^{n}}{M^{n}}
-\mu^{2}+2M^{2}+M^{2}\left(\frac{E_{1}}{E_{2}}+\frac{E_{2}}{E_{1}}\right)}
{2E_{1}E_{2}+2AE_{1}E_{2}\frac{E_{1}^{n}+E_{2}^{n}}{M^{n}}-M^{2}\left(\frac{E_{1}}{E_{2}}+\frac{E_{2}}{E_{1}}\right)}
+\mathcal{O}(A^{2})\nonumber\\
&>&1+\frac{2A\left(\frac{(E_{1}+E_{2})^{n+2}}{\mu^{n}}-\frac{E_{1}^{n+2}+E_{2}^{n+2}}{M^{n}}\right)-2A\frac{E_{1}^{n+1}E_{2}+E_{1}E_{2}^{n+1}}{M^{n}}}
{2E_{1}E_{2}+2A\frac{E_{1}^{n+1}E_{2}+E_{1}E_{2}^{n+1}}{M^{n}}}\nonumber\\
&>&1+\frac{2A}{M^{n}}\frac{\left(E_{1}+E_{2}\right)^{n+2}-E_{1}^{n+2}-E_{2}^{n+2}-E_{1}^{n+1}E_{2}-E_{1}E_{2}^{n+1}}
{2E_{1}E_{2}+2A\frac{E_{1}^{n+1}E_{2}+E_{1}E_{2}^{n+1}}{M^{n}}}\ ,
\end{eqnarray}
where the ultra relativistic approximation is involved (\(\mu\ll E,~M\ll E_{1},~M\ll E_{2}\)) in the third step.
It is easy to check that the right hand side of the above formula is always greater than \(1\).
Thus, the Cherenkov-like process is forbidden for the superluminal neutrinos in Finslerian special relativity
and the superluminal neutrinos would not lose their energy rapidly.

\section{III. Phenomenology: Energy dependence of the neutrino superluminality}

The superluminality was reported by the OPERA collaboration (OPERA) \cite{OPERA2011} to be
\(\delta v:=v-1=\left(2.18\pm0.77\pm0.30\right)\times10^{-5}\) for muon neutrinos with mean energy \(13.9~\rm{GeV}\)
and \(\delta v=\left(2.75\pm0.75\pm0.30\right)\times10^{-5}\) for muon neutrinos with mean energy \(42.9~\rm{GeV}\).
For all neutrinos with mean energy \(17~\rm{GeV}\), the superluminality is reported to be \(\delta v=(2.48\pm0.28\pm0.30)\times10^{-5}\).
The MINOS collaboration (MINOS) \cite{MINOS2007} reported that the superluminality is
\(\delta v=(5.1\pm 2.9)\times10^{-5}\) for muon neutrinos with \(3~\rm{GeV}\).
Report from the FermiLab in 1979 (FermiLab1979) \cite{Fermilab1979} showed that the muon neutrino with energy
between \(30~\rm{GeV}\) and \(120~\rm{GeV}\) may propagate superluminally with \(\delta v\sim10^{-5}\).
In addition, the observations of Supernova-1987A (SN1987A) \cite{SN1987A} set a stringent limit on
the supperluminal behaviors of antielectron neutrinos with energy \(\sim10~\rm{MeV}\) to be \(\delta v\lesssim2\times10^{-9}\).

As is mentioned in the introduction, the superluminality is revealed by the physical dispersion relations (\ref{Dispersion relations})
with extra terms which are dependent on the energy of particles phenomenally.
Especially, the simplest linear and quadratic energy dependence are considered popularly
which correspond to the five and six dimensional operators added to the neutrino Lagrangians
in the LIV models \cite{CacciapagliaDP001,EllisHMRS001,von GersdorffQ001}.
In addition, the data of OPERA and MINOS experiments revealed that the power exponent of energy dependence
should be in the range \(0.40-1.18\) \cite{Trojan001}.
However, the SN1987A observation showed that both linear and quadratic energy dependence are ruled out for
the neutrino superluminality \cite{AlexandreEM001}.
Only the energy dependence with higher orders than two could reconcile the datasets of SN1987A and OPERA experiments \cite{AlexandreEM001}.

In Finslerian special relativity, we have shown that the generic power-law dispersion relations (\ref{Dispersion relations})
are related to Finslerian structures leading to LIV.
In the following, we consider the possible energy dependence of the superluminal behaviors of neutrinos in the Finslerian framework.
The simple power-law energy dependence of neutrino superluminality is studied by combining the present observations of neutrino superluminality.
In addition, one of the simplest interpolations is considered to take account the stringent constraint on neutrino superluminality from the SN1987A.

\subsection{A. Energy independent superluminality}

The observed superluminality is reported to be at the order \(10^{-5}\) together with large errorbars for the muon neutrinos with
energy between \(\sim1~\rm{GeV}\) and \(\sim200~\rm{GeV}\) from the OPERA, MINOS and FermiLab1979 experiments \cite{OPERA2011,MINOS2007,Fermilab1979}.
In this energy range of neutrinos, the superluminality may be energy independent
\cite{CacciapagliaDP001,LiW001,Energy independent__Dass,Amelino-CameliaGLMRL001}
\begin{equation}
\delta v\sim\mathcal{O}\left(10^{-5}\right)\ ,
\end{equation}
which is consistent with the experimental datasets.
The energy independent superluminality of neutrinos corresponds to the physical dispersion relation
\begin{equation}
\label{energy independence}
\eta^{\mu\nu}p_{\mu}p_{\nu}=m^{2}-2Ap_{0}^{2}\ ,
\end{equation}
where the parameter \(A\sim\mathcal{O}\left(10^{-5}\right)\).
The dispersion relation has been proposed by the previous works on LIV at high energy scales \cite{SME,ColemanGlashowLIV}.
In addition, it is the dispersion relation (\ref{Finslerian special relativity dispersion relation}) with \(n=0\).

\begin{figure}[h]
\begin{center}
\includegraphics[width=8cm]{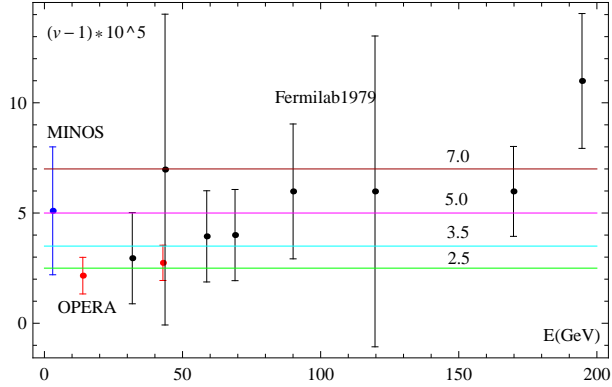}
\caption{\small{A schematic plot for the energy independence of neutrino superluminality.
The errorbars denote the datasets of observations on superluminal neutrinos from
OPERA (red) \cite{OPERA2011}, MINOS (blue) \cite{MINOS2007} and FermiLab1979 (black) \cite{Fermilab1979} experiments.
The horizontal lines denote the energy independent superluminal behaviors of neutrinos
which are set to be \(\delta v=(2.5,~3.5,~5.0,~7.0)\times10^{-5}\) from below to above.}}
\label{fig2}
\end{center}
\end{figure}

In Finslerian special relativity, the dispersion relation (\ref{energy independence}) is related to the Finslerian structure
\begin{equation}
F(y)=\alpha\left(1-A\left(\frac{\beta}{\alpha}\right)^{2}\right)\ .
\end{equation}
Meanwhile, the neutrino superluminality from the SN1987A is \(10^{4}\) times less than \(10^{-5}\) for
antielectron neutrinos with mean energy \(\sim10~\rm{MeV}\).
Thus, the energy threshold of superluminality may exist and should be much higher than \(\sim10~\rm{MeV}\) for neutrinos.
In other words, the LIV of superluminal neutrinos may be ``mass'' dependent \cite{LiW001}.
The effects of Finslerian structure may emerge and impact on the neutrino superluminality above this energy threshold.
The spacetime background may be modified by the Finsler geometry and
the Minkowski structure may be altered by the Finslerian structure above the huge Lorentz boosts related to this energy threshold for the neutrinos.
The energy independent superluminality of neutrinos is illustrated in Fig.\ref{fig2}.

\subsection{B. Linear energy dependent superluminality}

Previous studies \cite{Finsler special relativity,LiW001,Energy independent__Dass,Amelino-CameliaGLMRL001} showed that
the linear energy dependence of superluminality could also account for the superluminal behaviors of neutrinos
observed by the OPERA, MINOS and FermiLab1979 experiments because of the large errorbars of these experimental data.
In general, the linear energy dependence of neutrino superluminality could be revealed as
\begin{equation}
\delta v=aE+b\ ,
\end{equation}
where the parameter \(a=A/m\) in Finslerian special relativity and the parameter \(b\) denotes an offset term.
This superluminality corresponds to the dispersion relation
\begin{equation}
\eta^{\mu\nu}p_{\mu}p_{\nu}=m^{2}-2ap_{0}^{3}-2b p_{0}^{2}\ .
\end{equation}

\begin{figure}[h]
\begin{center}
\includegraphics[width=8cm]{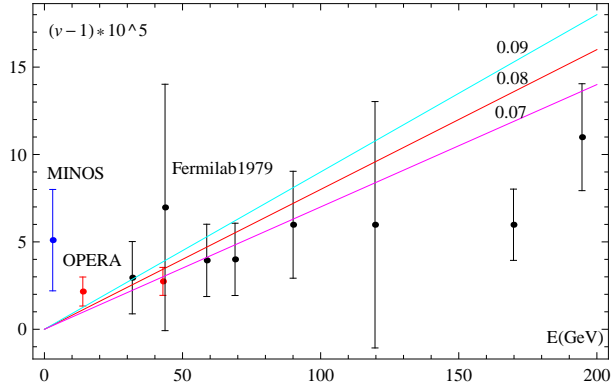}
\caption{\small{A schematic plot for the linear energy independence without offsets of neutrino superluminality.
The parameter $A$ is set to be \((7,~8,~9)\times10^{-18}\) from below to above and the neutrino mass is chosen to be \(0.01~\rm{eV}\).}}
\label{fig3}
\end{center}
\end{figure}

In the case that the offset term vanishes \(b=0\), the above neutrino experiments showed that
the parameter \(a^{-1}\sim10^{6}~\rm{GeV}\) is roughly consistent with the experimental data \cite{Finsler special relativity,LiW001}.
The corresponding Finslerian structure is given by \cite{Finsler special relativity}
\begin{equation}
F(y)=\alpha\left(1-A\left(\frac{\beta}{\alpha}\right)^{3}\right)\ .
\end{equation}
The upper limit of the muon neutrino mass is at the order \(0.01~\rm{eV}\) \cite{PDG}.
Thus, the parameter \(A\) is set to be of order \(10^{-17}\) in this case.
The linear energy dependence without offsets of the superluminal behaviors of neutrinos is illustrated in Fig.\ref{fig3}.

\begin{figure}[h]
\begin{center}
\includegraphics[width=8cm]{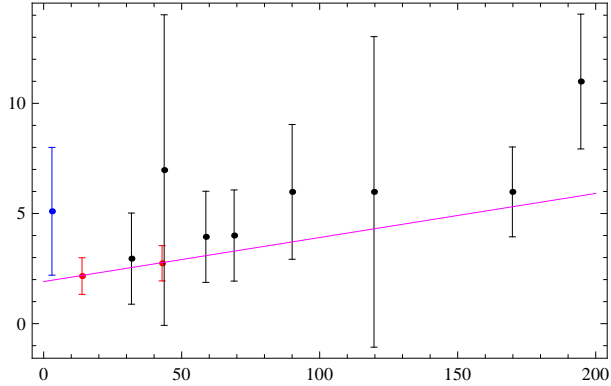}
\caption{\small{A schematic plot for the linear energy independence without offsets of neutrino superluminality.
The parameter $A$ is constrained to be \(2\times10^{-18}\) and the parameter $b$ is constrained to be \(1.91\times10^{-5}\) for the purplish red fit-line.}}
\label{fig4}
\end{center}
\end{figure}

In the case that the offset term exists \(b\neq0\), a very nice fit of the observed data is
given by \(a^{-1}=5\times10^{6}~\rm{GeV}\) and \(b=1.91\times10^{-5}\), namely \cite{Energy independent__Dass,Amelino-CameliaGLMRL001}
\begin{equation}
\delta v=2\times10^{-7}E_{GeV}+1.91\times10^{-5}\ ,
\end{equation}
where the lower index \(\rm{GeV}\) denotes the energy unit of neutrinos.
This superluminal case is related to the Finslerian structure of the form
\begin{equation}
F(y)=\alpha\left(1-A\left(\frac{\beta}{\alpha}\right)^{3}-b\left(\frac{\beta}{\alpha}\right)^{2}\right)\ ,
\end{equation}
where the parameter \(A\) is at the order \(2\times10^{-18}\).
It is noted that it is difficult to reconcile the datasets of OPERA and SN1987A in the simplest linear energy dependent scenario.
The energy threshold of neutrino superluminality may also appear in this linear scenario.
This linear energy dependence with offsets of neutrino superluminality could be illustrated in Fig.\ref{fig4}.

\subsection{C. Power-law energy dependent superluminality}

Both the energy independence and linear dependence of neutrino superluminality are consistent with
the datasets of present OPERA, MINOS and FermiLab1979 observations.
However, the data of SN1987A showed that the energy independent superluminality should disappear under certain energy threshold.
In addition, the linearly dependent superluminality of neutrinos is slightly larger than the observed upper limit \(2\times10^{-9}\)
although the superluminality is predicted to be at the same order \(10^{-9}\).
In the low energy ranges, the neutrino superluminality is suppressed to be smaller than that in the high energy ranges.
This fact may trigger the studies on the power-law dependence with higher orders which is even more steep than the linear case.

The nonlinear energy dependence of superluminality means power-law energy dependence with higher orders for the superluminal neutrinos.
To account for the data of SN1987A observations, the simplest power-law behaviors of superluminal neutrinos is considered popularly
\begin{equation}
\delta v=aE^{i}_{GeV}\ ,
\end{equation}
where the parameters \(a\) and \(i\) should be determined by the experimental observations.
This kind of neutrino superluminality corresponds to
the Finslerian line element (\ref{Finslerian special relativity line elements}) and
the dispersion relation (\ref{Finslerian special relativity dispersion relation}) with \(n=i\).
By combining the data of OPERA and MINOS, it is found that the dimensional parameter \(a\) should be in the range \((0.09-16.6)\times10^{-5}\)
and the dimensionless power exponent \(i\) should be within \(0.40-1.18\) \cite{Trojan001}.
However, it is argued that the data of SN1987A rules out the linear and quadratic dependence of the neutrino superluminality \cite{AlexandreEM001}.
More detailed discussions showed that the SN1987A data disfavors all \(i<2.5\) \cite{Amelino-CameliaGLMRL001}.

\subsection{D. Interpolations}

As is discussed in the last subsections, the OPERA's data requires flat energy dependence of neutrino superluminality
while the SN1987A's data requires more steep energy dependence of superluminality.
To reconcile the SN1987A and OPERA observations, it is essential to balance these two crosscurrents.
In principle, we could always make reconcilement between the datasets of superluminal neutrinos from the SN1987A and OPERA observations.
One of the simplest means to realize this purpose is to find an energy dependent function
so that the superluminal behaviors are steep at low energy ranges while they become flat at high energy ranges.
However, we have demonstrated that it is difficult to realize this purpose for the function with the simplest power-law energy dependence
in which there are equal or less than two parameters.
If more parameters are involved, it is possible to reconcile the datasets of SN1987A and OPERA observations.
For instance, the Lifshitz-type fermion model implies energy dependence of neutrino superluminality with even powers of high orders
to realize this purpose\cite{AlexandreEM001}.

\begin{figure}[h]
\begin{center}
\includegraphics[width=8cm]{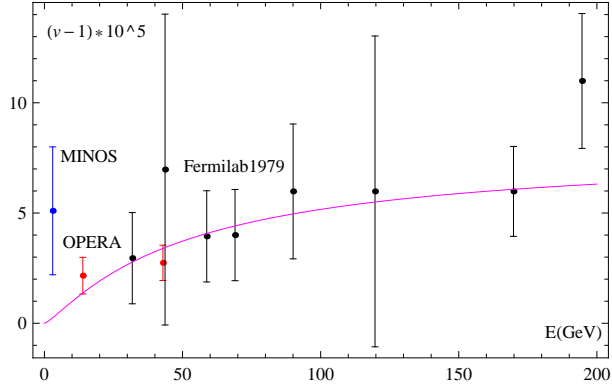}
\caption{\small{A schematic plot for the interpolation (\ref{interpolation}) of all datasets of superluminal behaviors of neutrinos from OPERA, MINOS and FermiLab1979 experiments. The purplish red curve is the interpolating curve for which the parameters are constrained to be \(a\approx8.0\times10^{-5}\), \(b^{-1}\approx40~\rm{GeV}\) and \(i\approx1.3\). For the neutrinos from SN1987A, the superluminality is consistent with the stringent upper limit \(2\times10^{-9}\) at the energy \(\sim10~\rm{MeV}\).}}
\label{fig5}
\end{center}
\end{figure}

The forms of interpolation are various to reconcile the present observed datasets of superluminal neutrinos.
It is unpractical to explore all the possible interpolations in this paper.
As an example, we consider one of the simplest interpolations of the form
\begin{equation}
\label{interpolation}
\delta v=a\left(\frac{bE}{bE+1}\right)^{i}\ ,
\end{equation}
where the parameters \(a\), \(b\) and \(i\) should be constrained by the present neutrino observations.
Here, one extra dimensional parameter \(b\) is involved, which denotes a typical energy scale of superluminal neutrinos.
The datasets of SN1987A, OPERA and FermiLab1979 observations constrain the parameters to be
\(a\approx8.0\times10^{-5}\), \(b^{-1}\approx40~\rm{GeV}\) and \(i\approx1.3\).
The superluminal curve determined by these parameters get through almost all errorbars of the neutrino superluminality
from these present territory observations.
The neutrino superluminality is less than \(2\times10^{-9}\) at the energy ranges related to neutrinos from SN1987A,
which is consistent with the astrophysical observation of SN1987A.
It is noted that this kind of superluminality could also be realized in the framework of Finslerian special relativity
since the formula (\ref{interpolation}) could be expanded into power-law series generally but with complex forms.
In addition, the above interpolation (\ref{interpolation}) could be illustrated in Fig.\ref{fig5}.

\section{IV. Discussions and Remarks}

The OPERA's report challenges cruelly the foundation of modern physics.
If it is confirmed in future, the neutrino superluminality would improve our knowledge on spacetime structure.
The superluminal behaviors of neutrinos may imply that the nature of spacetime is different from the Minkowski one
and the Lorentz symmetry should be replaced by some new symmetry.
In such a new spacetime, the neutrino superluminality is admitted and would not be suppressed by the Cherenkov-like process.
Meanwhile, the causal law is preserved and the energy-momentum is conserved.
Most importantly, the theoretical predictions on neutrino superluminality should be consistent with the experiments and observations.

In our previous paper (arXiv:1110.6673 [hep-ph]), we have proposed Finslerian special relativity as a reasonable candidate
to account for the OPERA neutrino superluminality.
Finslerian special relativity resides in the Finsler spacetime where the LIV is admitted.
It was found that Finslerian special relativity meets the above requirements of the new spacetime
and the linear energy dependence of superluminality is consistent with the data of the present observations on superluminal neutrinos.

In this paper, we investigated the symmetry, causal structure and superluminality in Finslerian special relativity.
In a generic case, the superluminal behaviors of neutrinos are of power-law energy dependence.
It was found that the generic Finslerian special relativity also admits the existence of neutrino superluminality
and the superluminal neutrinos would not lose their energy via the Cherenkov-like process rapidly.
Both the causality and energy-momentum conservation are preserved.
In addition, we studied the dispersion relations with extra power-law terms of higher orders corresponding to the Finslerian structures.
These dispersion relations were compared with the datasets of the present observations of superluminal neutrinos in detail.
It was found that Finslerian special relativity could be a reasonable arena to interpret the neutrino superluminality
at least in the energy ranges of present territory experiments and astrophysical observations.
Of course, more observable datasets on superluminal neutrinos are required to test and discriminate the models.
We wish that the MINOS experiment and the T2K experiment could give opportunities
to test and discriminate the predictions of Finslerian special relativity in future.

\begin{acknowledgments}

We thank useful discussions with Y. G. Jiang, M. H. Li and H. N. Lin.
This work is supported by the National Natural Science Fund of China under Grant No. 10875129 and No. 11075166.

\end{acknowledgments}

\end{document}